\documentclass[11pt]{revtex4}
\usepackage{amssymb,epsf}
\usepackage{epsfig}
\usepackage{graphicx, latexsym, amssymb, amscd, psfrag}
\usepackage{latexsym}
\begin{document}

\title{Static Properties of the Multiple-Sine-Gordon Systems}
\author{M. Peyravi$^{1}$, N. Riazi$^{1}$\footnote{email: riazi@physics.susc.ac.ir}and Afshin Montakhab$^{2}$\footnote{email: montakhab@shirazu.ac.ir}
} \affiliation{$1$. Physics Department and Biruni Observatory, Shiraz University, Shiraz 71454, Iran,\\
$2$. Physics Department, Shiraz University, Shiraz 71454, Iran. }

\begin{abstract}
In this paper, we examine some basic properties of the
multiple-Sine-Gordon (MSG) systems, which constitute a
generalization of the celebrated sine-Gordon (SG) system. We start
by showing how MSG systems can be viewed as a general class of
periodic functions. Next, periodic and step-like solutions of
these systems are discussed in some details. In particular, we
study the static properties of such systems by considering slope
and phase diagrams. We also use concepts like energy density and
pressure to characterize and distinguish such solutions. We
interpret these solutions as an interacting many body system, in
which kinks and antikinks behave as extended particles. Finally,
we provide a linear stability analysis of periodic solutions which
indicates short wavelength solutions to be stable.
\end{abstract}

\maketitle
\section{Introduction\label{intro}}

The double-Sine-Gordon (DSG) equation which is a generalization of
the ordinary Sine-Gordon (SG) equation has been the focus of much
recent investigations \cite{0,1,2,3,4,5,6,7,8,9,10,11,12,13,14}.
It has been shown to model a variety of systems in condensed
matter, quantum optics, and particle physics \cite{1}. Condensed
matter applications include the spin dynamics of superfluid $^3He$
\cite{2,3}, magnetic chains \cite{4}, commensurate-incommensurate
phase transitions \cite{5}, surface structural reconstructions
\cite{6}, and domain walls \cite{7,8} and fluxon dynamics in
Josephson junction \cite{9}.

\begin{figure}
\epsfxsize=10cm \centerline{\epsfbox{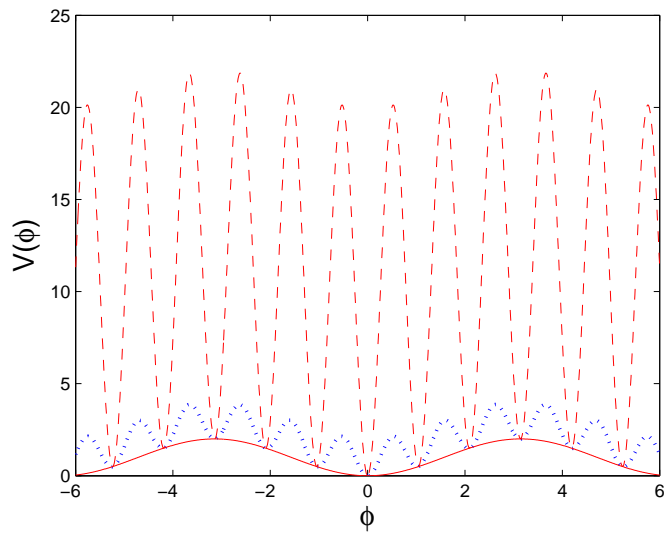}} \caption{MSG
Potential for $N=6$. The dashed curve is for $\epsilon=10$, the
dotted curve is for $\epsilon=1$ and the solid curve is for
$\epsilon=0$.\label{1}}
\end{figure}

 In quantum field theory and quantum
optics, DSG applications include quark confinement \cite{10} and
self-induced transparency \cite{11}. The internal dynamics of
multiple and single DSG soliton configurations using molecular
dynamics have been studied in \cite{1}. There have also been
studies about kink anti-kink collision processes for DSG equation
\cite{12}. One can also point to the statistical mechanics
applications \cite{13}, and perturbation theory for this system
\cite{14}. Following our pervious study on the periodic and
step-like solutions of DSG equation\cite{0}, we focus on a
generalization of this system with the self-interaction potential
(see Fig.\ref{1})
\begin{equation}
V(\phi)=1+\epsilon -\cos \phi -\epsilon \cos (N\phi ),
\end{equation}
where $\epsilon$ is a non-negative constant. This potential
reduces to the ordinary SG potential in the limit
$\epsilon\rightarrow 0$. $N$ is an integer with $N=1$ and  $N=2$
corresponding to the SG and  DSG systems, respectively
\cite{0,ri1}. In this paper, we propose to study this system along
the line of our previous work on DSG \cite{0} for $N>2$. In
Section \ref{sec2} we present some general properties of the
system including the single kink solutions, in Section \ref{sec3}
we examine the $N=3$ case, in Section \ref{sec4}, $N\geq4$ cases
are discussed, and in Section \ref{sec5} the stability of periodic
solutions is examined. Section \ref{sec6} is devoted to a summary
and conclusions.

\section{General properties of the MSG system}\label{sec2}

The Lagrangian density of the Multiple-Sine-Gordon (MSG) system is
the following:
\begin{equation}
{\cal L}_{MSG}=\frac{1}{2} \partial^\mu \phi \partial_\mu \phi
-\left[ 1+\epsilon -\cos\phi -\epsilon \cos (N\phi
)\right].\label{msg}
\end{equation}
From this Lagrangian density, the MSG equation follows
\begin{equation}
\Box \phi =-\sin \phi -N\epsilon \sin (N\phi ).\label{shok}
\end{equation}
Note that $\phi$ is a real scalar field, and we are using a 1+1
dimensional spacetime (t,x) with the signature ($+$,$-$). The
potential for this system has minima at
\begin{equation}
\phi_{min}=2n\pi  \qquad \ n=0,1,2,...
\end{equation}

\begin{figure}[h]
\epsfxsize=8.7cm\centerline{\hspace{8cm}\epsfbox{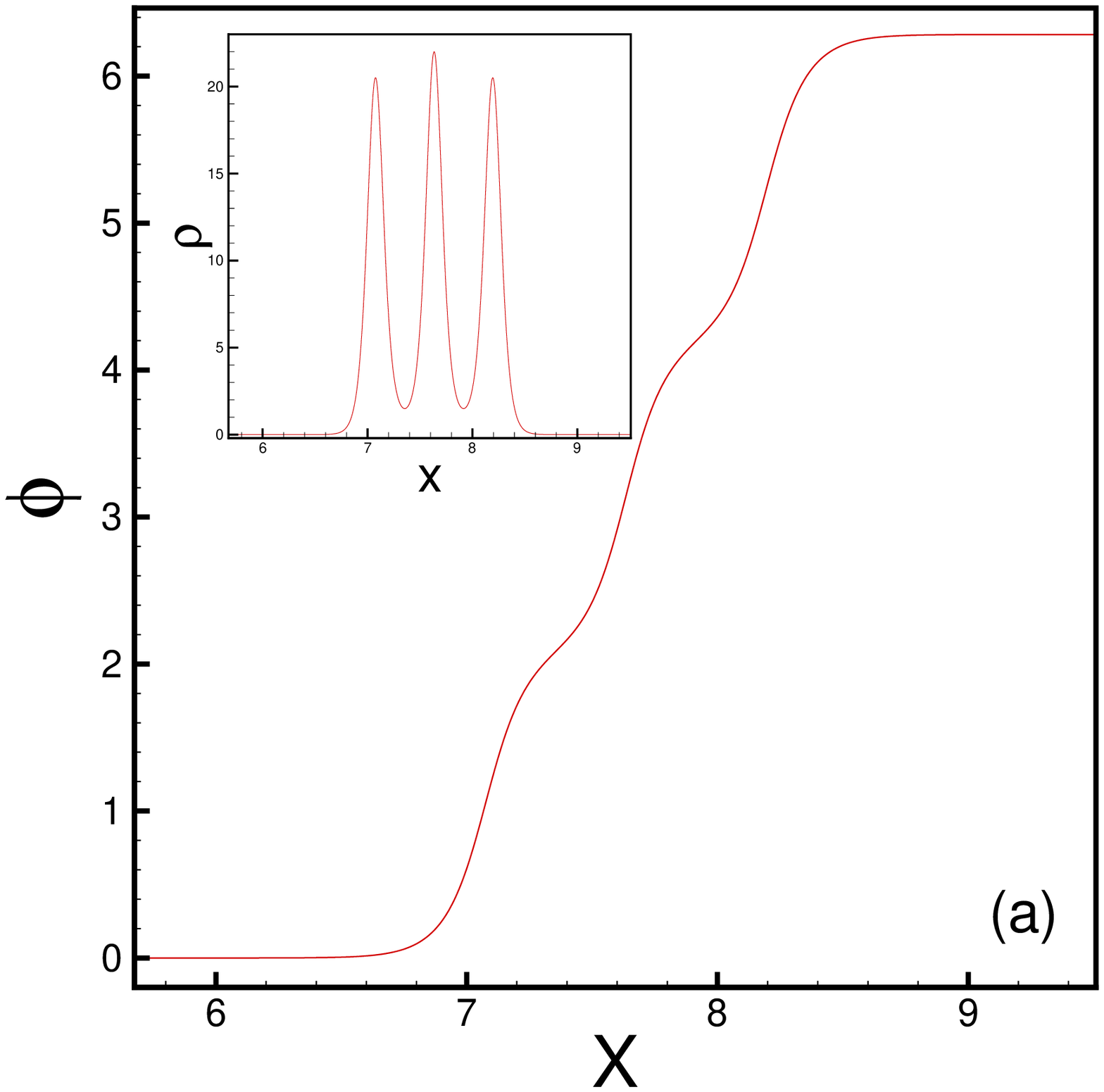}\epsfxsize=9cm\centerline{\hspace{-7cm}\epsfbox{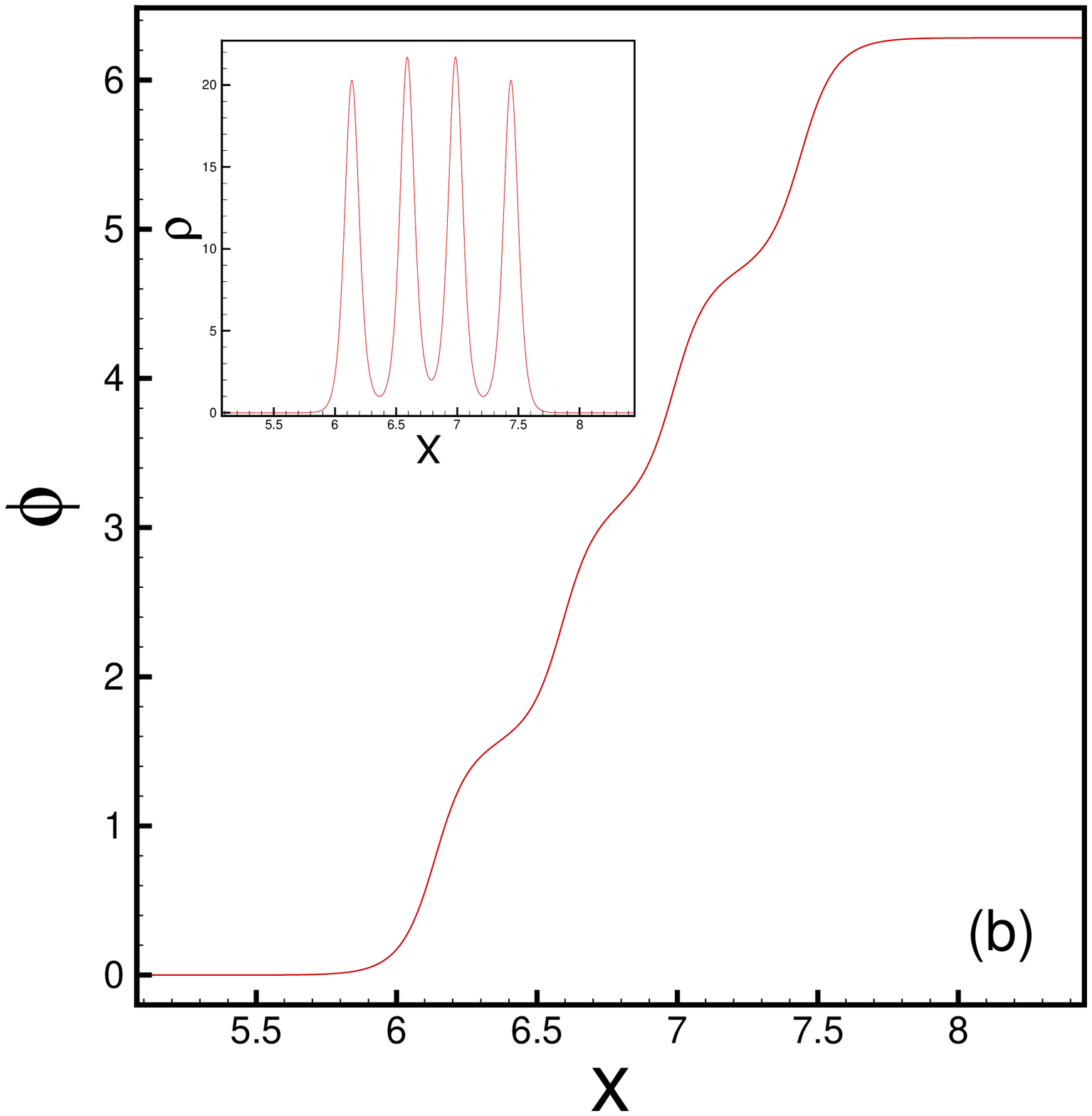}}}
\caption{Single soliton solutions and energy densities (inset) of
the MSG system for (a)$N=3$, $\epsilon=10$ and (b)$N=4$,
$\epsilon=10$.\label{2}}
\end{figure}

The energy-momentum tensor of the MSG equation can be obtained by
the standard relation which follows from the Noether's theorem and
is a consequence of the translational invariance of the Lagrangian
density:
\begin{equation}
T^{\mu\nu}=\partial^\mu \phi \partial^\nu \phi -g^{\mu\nu}{\cal
L}_{MSG},
\end{equation}
in which $g^{\mu\nu}=diag(1,-1)$ is the metric of the Minkowski
spacetime.

Like the ordinary SG system, a topological current can also be
defined for the MSG system according to
\begin{equation}
J^\mu =\frac{1}{2\pi}\epsilon^{\mu\nu}\partial_\nu \phi,
\end{equation}
where $\epsilon^{\mu\nu}$ is the totally antisymmetric tensor with
$\epsilon^{01}=1$.  This current is identically conserved
($\partial_{\mu}J^{\mu}=0$) and the total charge of any localized,
finite-energy solution is both conserved and quantized.

MSG equation possesses kink and anti-kink solutions which
correspond to transitions between the spatial boundary conditions
$\phi (\pm \infty )=2n\pi$. The first integral of equation
(\ref{shok}) for static solutions reads
\begin{equation}
\frac{1}{2}(\frac{d\phi}{dx})^2 =V(\phi),\label{kinkp}
\end{equation}
in which we have used the boundary conditions $\phi (\pm \infty
)=2n\pi$. We therefore have
\begin{equation}
x-x_0=\int \frac{d\phi}{\sqrt{2V(\phi )}}.
\end{equation}
This integral cannot be carried out analytically for general $N$
and $\epsilon$. A few examples of these solutions for $N=3$ and
$N=4$ which are obtained numerically are shown in Fig.\ref{2},
together with their energy density, which is equal to the
$T^{0}_{0}$ component of $T^{\mu}_{\nu}$.

Multiple Sine-Gordon systems admit soliton-like kink solutions
with interesting properties. Here, we show that an arbitrary,
periodic potential with vanishing minima at $\phi=0$ and
$\phi=2\pi$ can be expanded in terms of MSG potentials. This
property adds to our motivation for studying this class of
nonlinear equations.

Let us define the MSG basis functions
\begin{equation}
F^{\epsilon}_{N}(\phi)\equiv1+\epsilon-\cos\phi-\epsilon\cos(N\phi).\label{f1}
\end{equation}
An arbitrary, periodic function $f(\phi)$ which satisfies the
boundary conditions $f(0)=f(2\pi)=0$ can be expanded in terms of
this basis:
\begin{equation}
f(\phi)=\sum_{N=0}^{\infty}a_{N}F^{\epsilon}_{N}(\phi)=(1+\epsilon)\sum_{N=0}^{\infty}a_{N}-\cos\phi(\sum_{N=0}^{\infty}a_{N})-
\epsilon\sum_{N=0}^{\infty}a_{N}\cos(N\phi),\label{f2}
\end{equation}
in which $a_{N}$'s are the expansion coefficients. At the same
time, $f(\phi)$ can also be represented by a cosine Fourier
series:
\begin{equation}
f(\phi)=\sum_{N=0}^{\infty}b_{N}\cos(N\phi).\label{f3}
\end{equation}
of course, $f(0)=f(2\pi)=0$ demands that
$\sum_{N=0}^{\infty}b_{N}=0.$

 Equating the coefficients of
$\cos(N\phi)$ in (\ref{f2}) and (\ref{f3}) leads to
\begin{eqnarray}
b_{0}&=&(1+\epsilon)\sum_{N=0}^{\infty}a_{N}-\epsilon
a_{0},\nonumber\\
b_{1}&=&-\sum_{N=0}^{\infty}a_{N}-\epsilon
a_{1},\nonumber\\
b_{N}&=&-\epsilon a_{N}  \qquad   (N>1).\label{f4}
\end{eqnarray}

Note that $\sum_{N=0}^{\infty}a_{N}$ can be easily calculated via
\begin{equation}
\sum_{N=0}^{\infty}a_{N}=\frac{1}{2\pi(1+\epsilon)}\int_{0}^{2\pi}
f(\phi)d\phi.\label{f5}
\end{equation}

As an example, consider the triangular periodic function
\begin{displaymath}
f(\phi)=\left \{\begin{array}{ccccc}
 \phi  \qquad 0\leqslant\phi\leqslant\pi\\
  2\pi-\phi \qquad \pi\leqslant\phi\leqslant2\pi\\
\end{array} \right.
\end{displaymath}
with $f(\phi+2n\pi)=f(\phi)$. The Fourier cosine series
representing this function reads
\begin{equation}
f(\phi)=\pi-\frac{8}{\pi}\cos(\phi)-\frac{8}{9\pi}\cos(3\phi)-\frac{1}{4\pi}\cos(4\phi)-....
\end{equation}

The coefficients $a_{N}$ can be easily calculated from Equations
(\ref{f4}) and (\ref{f5}):
\begin{eqnarray}
\sum_{N=0}^{\infty}a_{N}&=&\frac{\pi}{2(1+\epsilon)}, \nonumber\\
a_{0}&=&-\frac{\pi}{2\epsilon}, \nonumber\\
a_{1}&=&\frac{16(1+\epsilon)-\pi^{2}}{2\pi\epsilon(1+\epsilon)},\nonumber\\
a_{2}&=&0, \nonumber\\
a_{3}&=&\frac{8}{9\pi\epsilon},\nonumber\\
a_{4}&=&\frac{1}{4\pi\epsilon},\nonumber\\
...\textrm{etc}
\end{eqnarray}

The reader may get worried about the $\frac{1}{\epsilon}$
dependence as $\epsilon\longrightarrow\infty$, but this is not a
serious problem, since this is a common factor for all $a_{N}$'s
and may be factored out.

\begin{figure}[h]
\epsfxsize=8.5cm\centerline{\hspace{8cm}\epsfbox{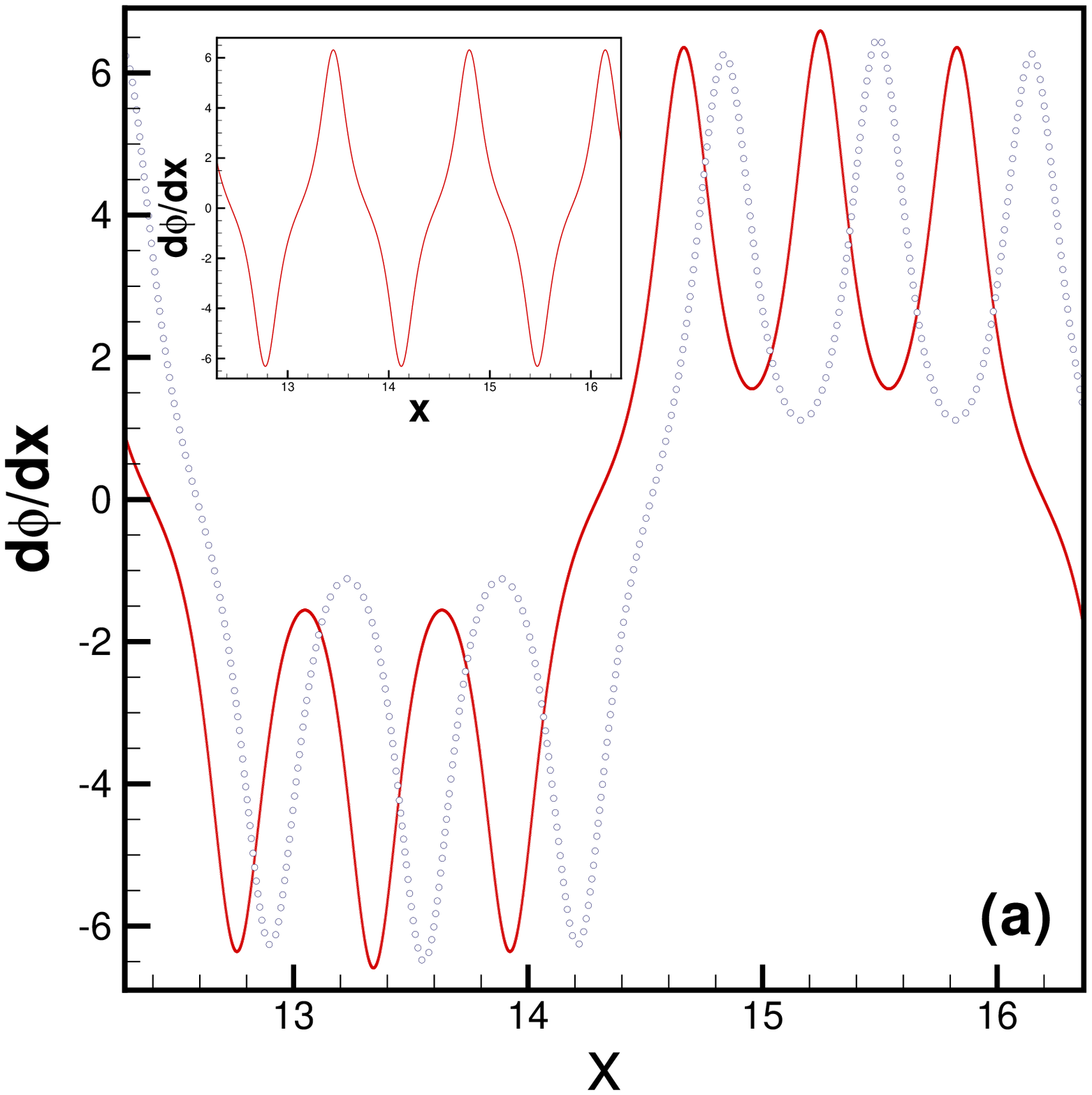}\epsfxsize=8.5cm\centerline{\hspace{-7cm}\epsfbox{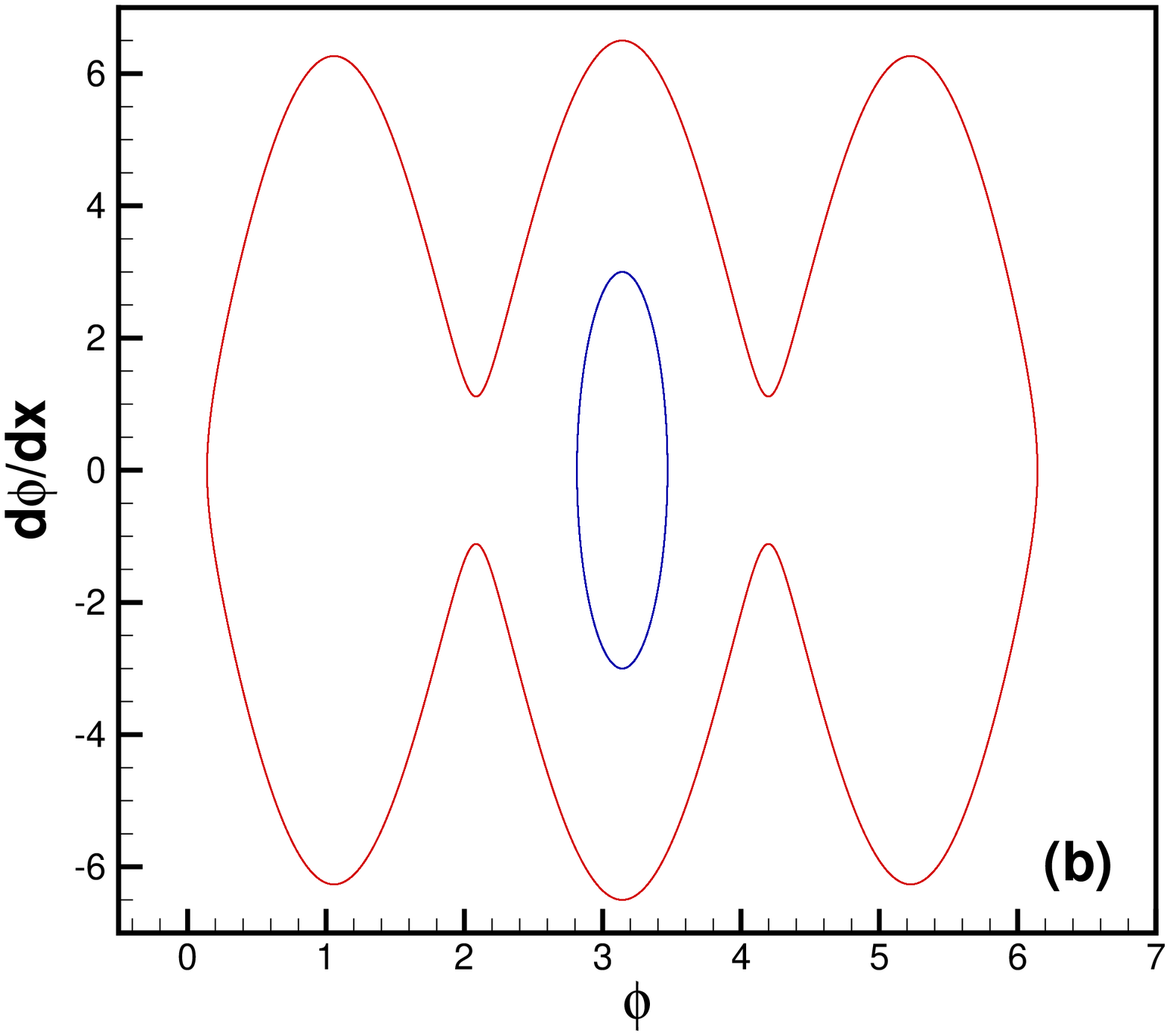}}}\caption{
(a) The slope diagram for the periodic chain of MSG  solitons for
$N=3$ and $\epsilon=10$, with $P= -0.2859 $ for the solid curve
and with $P=-0.8750$ dotted curve and  $P=-2.0604 $ for the inset.
The three solutions have the same average density ($\bar{\rho}$)
but different pressure ($P$). The main figure shows three subkink
solutions while the inset shows a no-subkink  solution. (b) The
phase diagram($\frac{d\phi}{dx}$ vs. $\phi$) for the periodic
chain of MSG solitons, for $N=3$ and $\epsilon=10$. The single
loop corresponds to $P=-17.5$, (kink anti-kink), while the three
lobe one corresponds to $P=-0.875$, (three subkinks).\label{3}}
\end{figure}

\section{MSG for $N=3$}\label{sec3}

In this and the next two sections, we investigate some general
properties of MSG for $N=3$, and subsequently for $N=4$ and $N>4$,
along the line of our previous work on DSG\cite{0}. We focus on
the static, time-independent solution of MSG equation. To solve
this equation, we use a fourth order Runge-Kutta method\cite{15}
to numerically integrate:
\begin{equation}
\frac{d^{2}\phi}{dx^{2}}=\frac{dV(\phi)}{d\phi}=\sin\phi+N\epsilon\sin(N\phi).
\end{equation}
 We use $\phi=\pi$ with various values of
$\frac{d\phi}{dx}$ at $x=0$, as our initial conditions throughout.
We study this system for various values of $N$ and $\epsilon$. The
solutions are characterized by the first integral of the static
MSG equation, i.e.
\begin{equation}
P=-T^{1}_{1}=\frac{1}{2}(\frac{d\phi}{dx})^2-V(\phi ),
\label{press}
\end{equation}
with
\begin{equation}
\frac{dP}{dx}=0.
\end{equation}
Note that $P$ is different from energy density
\begin{equation}
\rho=T^{0}_{0}=\frac{1}{2}(\frac{d\phi}{dx})^2+V(\phi);
\end{equation}
which changes with position $x$. The reason for using the letter
$P$ for this quantity is that it conforms with ``pressure'' for a
one dimensional perfect fluid:
\begin{equation}
T^\mu_\nu=diag(\rho,-P).
\end{equation}
Here, pressure is to be interpreted as tension, rather than force
per unit area \cite{0}. The value of $P$ distinguishes two
different type of solutions: (i) step-like solutions for which
$P>0$, which are the ``running'' solution of the mechanical
analog, (ii) periodic solutions for which $-2(1+\epsilon)<P<0$,
which are the ``oscillatory'' solutions of the mechanical
analog\cite{16}. Here, since we have chosen $\phi=\pi$ for all our
solutions  as an initial condition, the minimum value of $P$ for
periodic solutions is $-2$ for $N$ even, where this lower bound is
$\epsilon$-dependent for odd $N$.

\begin{figure}
\epsfxsize=10cm\centerline{\epsfbox{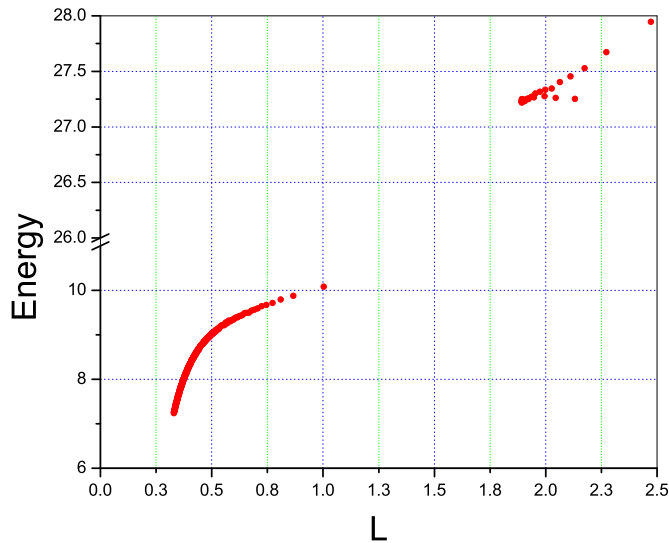}} \caption{Energy per
soliton diagram for periodic chain of MSG solitons for $N=3$ and
$\epsilon=10$. The low energy portion contains no subkinks while
the high energy solitons is multi-valued and contains
three-subkink solitons.\label{4}}
\end{figure}

We first consider the simpler, step-like solutions. These
solutions simply appear as in the case of DSG equations. The
general MSG equation exhibits step-like solutions for $P>0$ which
are always characterized by $N$ subkinks, as in Fig.\ref{2},
provided that $\epsilon$ is in the appropriate range. However, as
in the case of DSG equation, the periodic solutions exhibit a rich
set of behavior depending on the value of $P$ and $\epsilon$, as
well as $N$. A simple way to imagine these solutions is to
consider slope diagrams ($\frac{d\phi}{dx}$ vs. $x$) or phase
diagrams ($\frac{d\phi}{dx}$ vs. $\phi$). Such a set is shown in
Fig.\ref{3} for $N=3$ and $\epsilon=10$. We note that for $N=3$, a
kink anti-kink solution (with no subkinks) as well as a
three-subkink solution is possible. Part(a) shows the kink
anti-kink solution as an inset and two different three-subkinks in
the main panel. Part (b) shows the phase diagram where each
(subkink) solution is represented by a ``loop'' which is a helpful
way of distinguishing these solutions.

\begin{figure}
\epsfxsize=10cm\centerline{\epsfbox{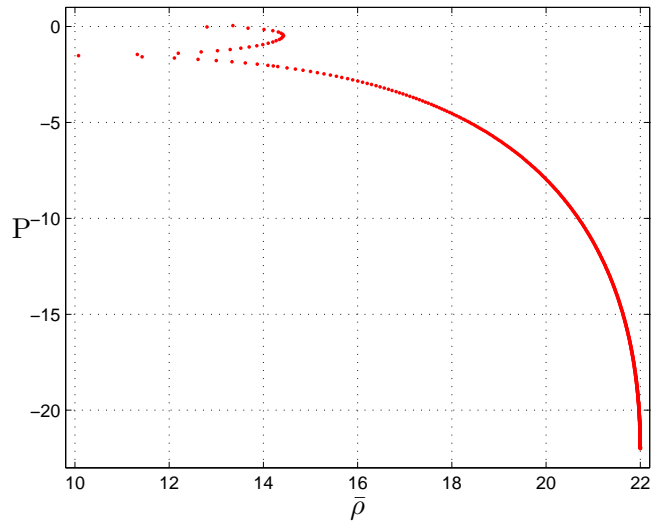}}\caption{Equation of
state diagram for periodic chain of MSG solitons for $N=3$ and
$\epsilon=10$. The upper curve corresponds to the three-subkink
solutions and lower one to the no-subkink solutions.\label{5}}
\end{figure}

\begin{figure}
\epsfxsize=10cm\centerline{\epsfbox{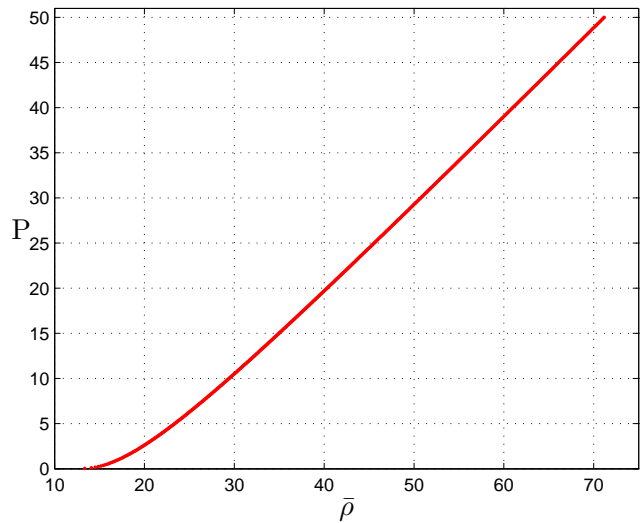}} \caption{The equation
of state diagram for N=3, $\epsilon=10$, for step-like solutions.
The $P\rightarrow 0$ limit corresponds to the single kink
solution. Unlike the periodic solutions, the general shape of this
curve is independent of $N$.\label{6}}
\end{figure}

A different (``thermodynamic'') method to characterize solutions
of such systems is to consider the energy diagram ($E$ vs. $L$) as
well as equation of state diagram ($P$ vs. $\bar{\rho}$)\cite{0}.
In the first case energy per soliton ($E$) is plotted as a
function of inter-soliton distance ($L$)\cite{17}, while in the
second case, $P$ is plotted as a function of average energy
density $\bar{\rho}=\frac{1}{l}\int_{0}^{l} \rho dx$ with $l$ is a
sufficiently large length (includes several solitons). These
figures are shown for $N=3$ and $\epsilon=10$ in Fig.\ref{4} and
Fig.\ref{5}, respectively. As in the case with DSG equation, the
energy diagram for $N=3$ shows a region where $E$ is
double-valued. That is, Fig.\ref{4} shows a low energy solution
with no subkinks, a region where no solutions are observed for our
initial conditions, and a high energy region which is double
valued. It is interesting to note that in the lower branch of high
energy solution, as $L$ increases, kink and anti-kink separate
out, while in the upper branch it is the subkinks (within a given
kink)which separate out when $L$ increases. We note that in this
three-subkink solution the middle subkink never changes and it is
the other two (first and last) which actually extend out in the
upper branch.

Fig.\ref{5} shows the equation of state for $N=3$ and
$\epsilon=10$. It consists of two separate branches as well. The
upper smaller branch is the three subkink solution and the lower
longer branch is the no subkink solution. The upper branch is
similar to that of DSG solution and the lower one is similar to
the regular SG equation. The tension (or negative pressure)
decreases in this 1d chain with increasing density due to the
attractive force between the solitons in the chain. The only
region where tension increases with increasing density is in the
lower half of the upper branch which indicates a different type of
inter-soliton force in the chain. It is worth noting that the
equation of state diagram for the step-like solutions for $N>2$ is
in general no different than the $N\leq2$ case previously studied
\cite{0}. Here we plot the $N=3$ case in Fig.\ref{6}.

\begin{figure}
\epsfxsize=7.2cm\centerline{\hspace{8cm}\epsfbox{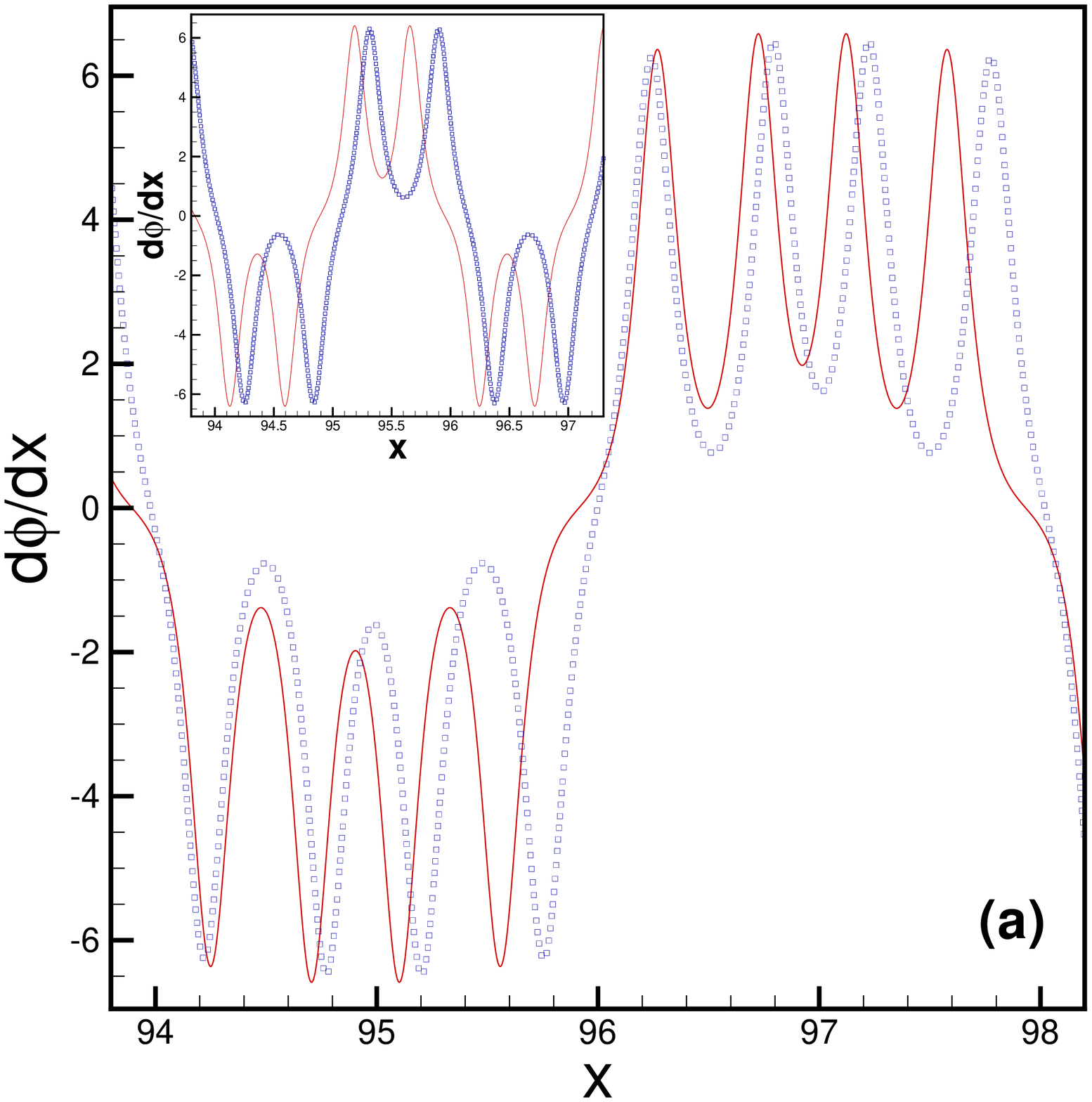}\epsfxsize=9.2cm\centerline{\hspace{-7cm}\epsfbox{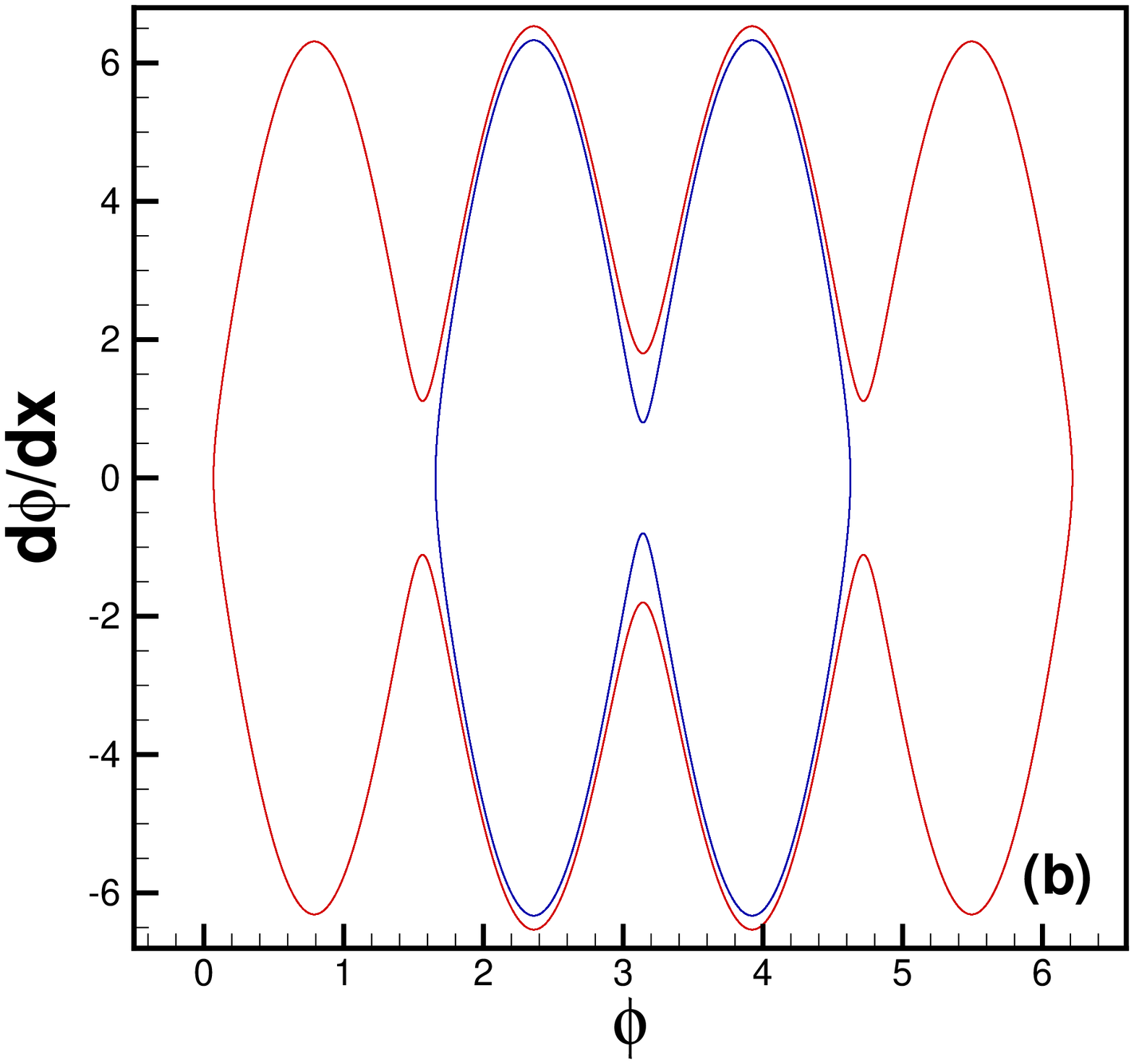}}}\caption{
(a) The slope diagram for the periodic chain of MSG  solitons for
$N=4$ and $\epsilon=10$, with $P= -0.0398 $ for the solid curve
and with $P= -0.70395$ dotted curve, with inset $P=-1.18080$ for
the solid curve and  $P=-1.8078$ dotted curve. The four solutions
have the same average density ($\bar{\rho}$) but different
pressure ($P$). The main figure shows four subkink solutions while
the inset shows two subkink solutions. (b) The phase
($\frac{d\phi}{dx}$ vs. $\phi$) diagram for the periodic chain of
MSG solitons for $N=4$ and $\epsilon=10$. The two lobe loop
corresponds to $P=-1.68$, while the four lobe one corresponds to
$P=-0.38$.\label{7}}
\end{figure}

\begin{figure}
\epsfxsize=10cm\centerline{\epsfbox{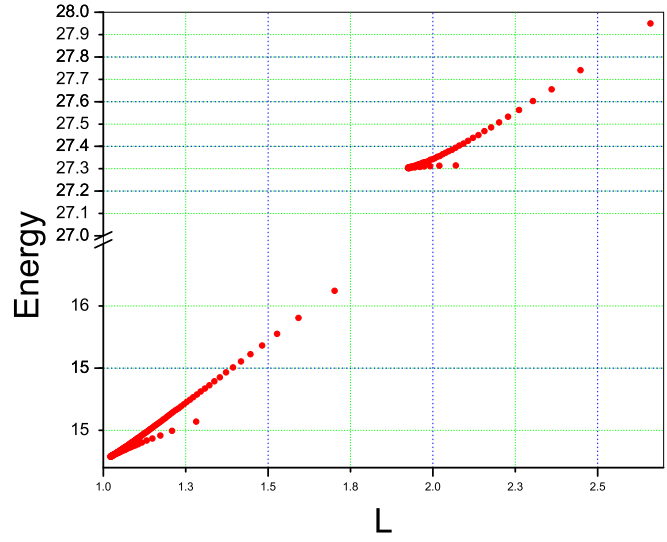}}\caption{Energy per
soliton diagram for periodic chain of MSG solitons for $N=4$ and
$\epsilon=10$. The upper curve corresponds to the four subkink
solutions and the lower one to the two subkink
solutions.\label{8}}
\end{figure}

\begin{figure}
\epsfxsize=10cm\centerline{\epsfbox{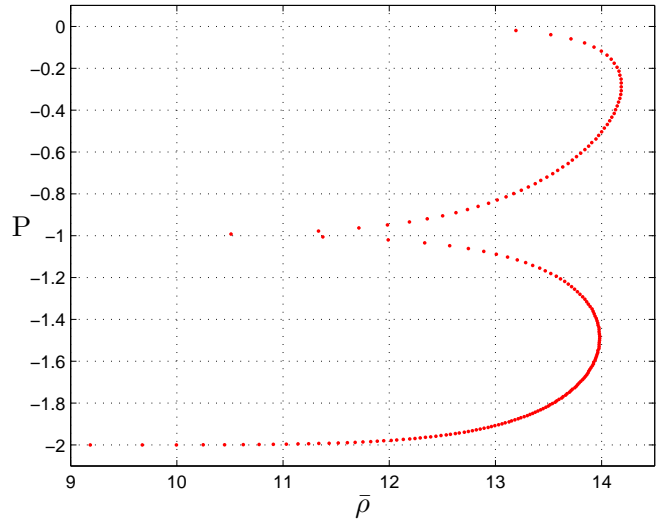}} \caption{The equation
of state diagram for $N=4$ and $\epsilon=10$ system. The upper
curve corresponds to the four subkink solutions and the lower one
to the two subkink solutions.\label{9}}
\end{figure}

\section{$N\geq4$ cases}\label{sec4}

\begin{figure}
\epsfxsize=20cm\centerline{\epsfbox{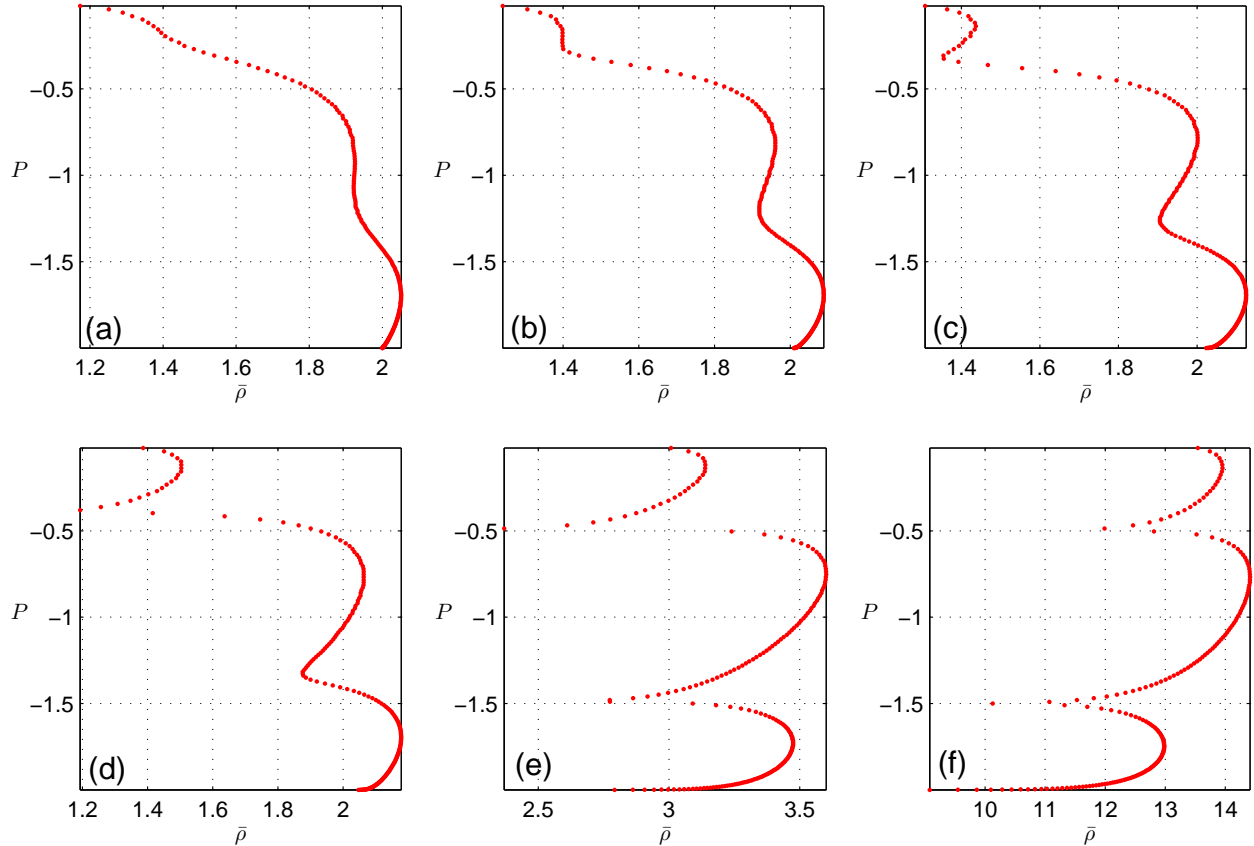}}\caption{The equation
of state diagram for $N=6$ and (a) $\epsilon=0.03$, (b)
$\epsilon=0.05$, (c) $\epsilon=0.07$, (d) $\epsilon=0.1$, (e)
$\epsilon=1$ and (f) $\epsilon=10$ system.\label{10}}
\end{figure}

\begin{figure}
\epsfxsize=10cm\centerline{\epsfbox{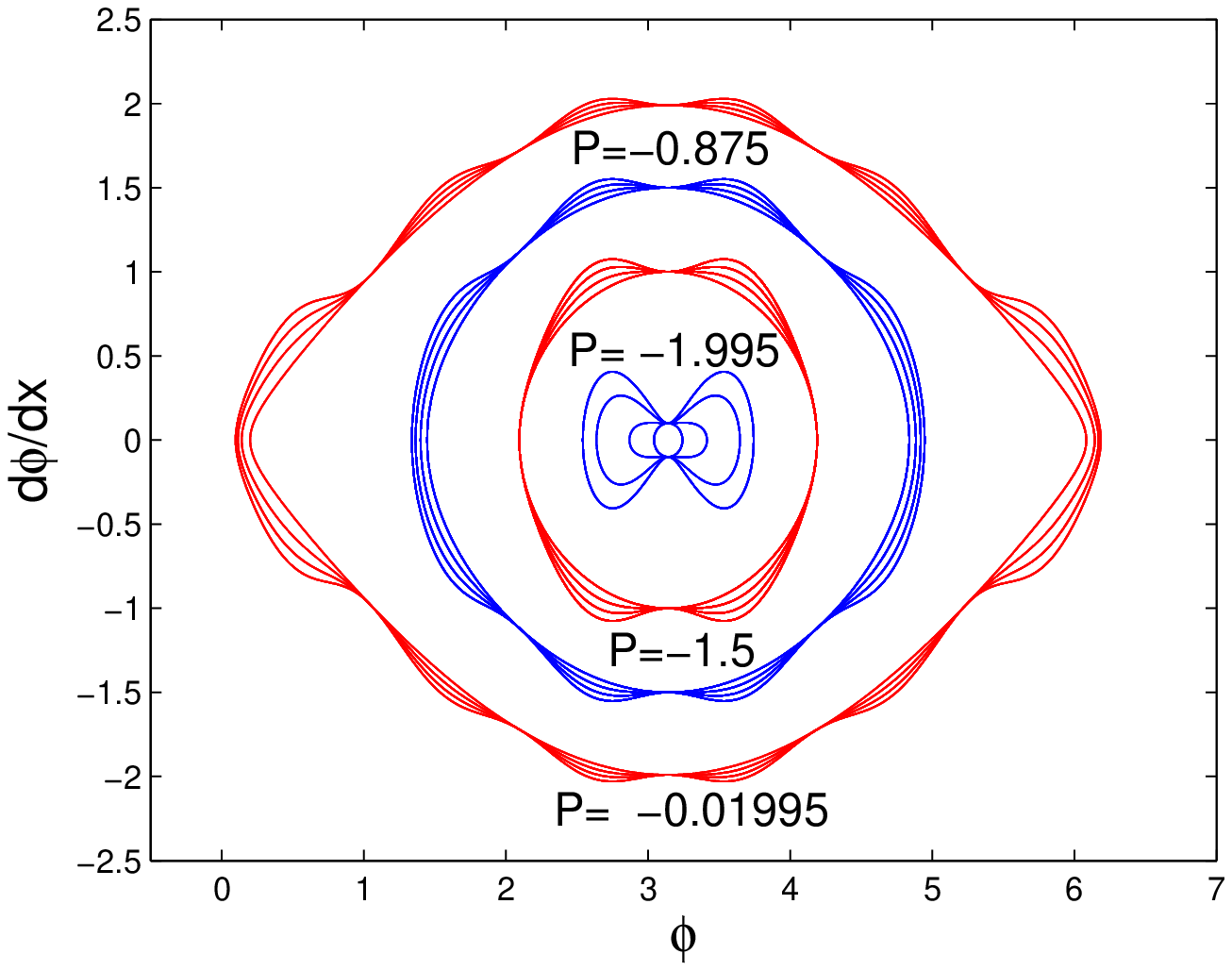}}\caption{The phase
($\frac{d\phi}{dx}$ vs. $\phi$) diagram for the periodic chain of
MSG solitons, for $N=6$. The nearby curves correspond to
$\epsilon=0, 0.03, 0.06$ and $0.09$ from the inner to the outer
ones. The inner set of loops corresponds to $P=-1.995$, as
indicated, all the way to $P=-0.01995$ for the other set of loops.
\label{11}}
\end{figure}

As $N$ increases, the MSG system gets more complicated and
detailed. The main reason for this is the development of more
subkinks. The emergence of subkinks in the periodic regime depends
on the value of $\epsilon$ as well as $P$. The maximum number of
possible subkinks is $N$. Due to the symmetry $V(\phi)=V(-\phi)$,
the number of subkinks (n) is also ``doubled'', i.e.
\begin{displaymath}
n=\left \{\begin{array}{ccccc}
  0 & 2 & 4 & ... & \textrm{up to $N$  \qquad for $N$ even}\\
  0 & 3 & 5 & ... & \textrm{up to $N$  \qquad for $N$ odd}\\
\end{array} \right.
\end{displaymath}

One can easily draw similar diagrams for general $N$ as the ones
draw for $N=3$ in the previous section. Here, we simply show these
in  Figs.\ref{7}-\ref{9}, for $N=4$. Here, two and four subkink
solutions are shown, with two subkink solutions being the low
energy solutions and thus lower pressure.

The change in the equation of state diagram as $\epsilon$
increases shows an interesting behavior. This is shown in
Fig.\ref{10} for $N=6$. It is interesting to note how the
multi-valuedness arises as $\epsilon$ is increased. As $\epsilon$
increases a larger range of $\bar{\rho}$ values allows multiple
solutions. Another interesting behavior is the emergence of cusp
points with a discontinuity in $\frac{dP}{d\bar{\rho}}$ (the left
most section in Fig.\ref{10} (e) and (f)) which is due to the
existence of hyperbolic unstable point in MSG equation. This
discontinuity dose not exist for the simple SG system. Note that
these cusps separate regions of solutions with different number of
subkinks. For example, in Fig.\ref{10}(f) two cusps separate the
6, 4 and 2 subkink solutions. Another interesting point in these
solutions is the maximum density solutions where compressibility
($\chi=\frac{d\bar{\rho}}{dP}$) becomes zero. At these point the
system becomes incompressible due to strong interactions of
kink-antikink solutions (one can see that this corresponds to a
minimum $L$)\cite{0}.

We are  also interested to see how different subkink solutions
arise as $\epsilon$ and $P$ are changed. We therefore plot the
phase diagram for various values of $P$ and look for the emergence
of subkinks as $\epsilon$ is increased. Such a plot for $N=6$ is
shown in Fig.\ref{11}. Here, each set of loops corresponds to a
given $P$ value. It is seen that with a small change in
$\epsilon$ subkink solutions arise. However, the number of
subkinks is more sensitive to change in $P$ than $\epsilon$. We
have observed that for a given value of $P$, increasing $\epsilon$
from zero can sometimes change the number of subkinks in a short
interval. This interval is typically of order of
$\epsilon\simeq0.1$. After this brief transition, the number of
subkinks remains constant. On the other hand the number of subkink
changes with increasing $P$ (for a given $\epsilon$) until it
reaches its maximum ($N$) for $P\longrightarrow0$.

\section{Stability around fixed points}\label{sec5}

\begin{figure}
\epsfxsize=10cm\centerline{\hspace{8cm}\epsfbox{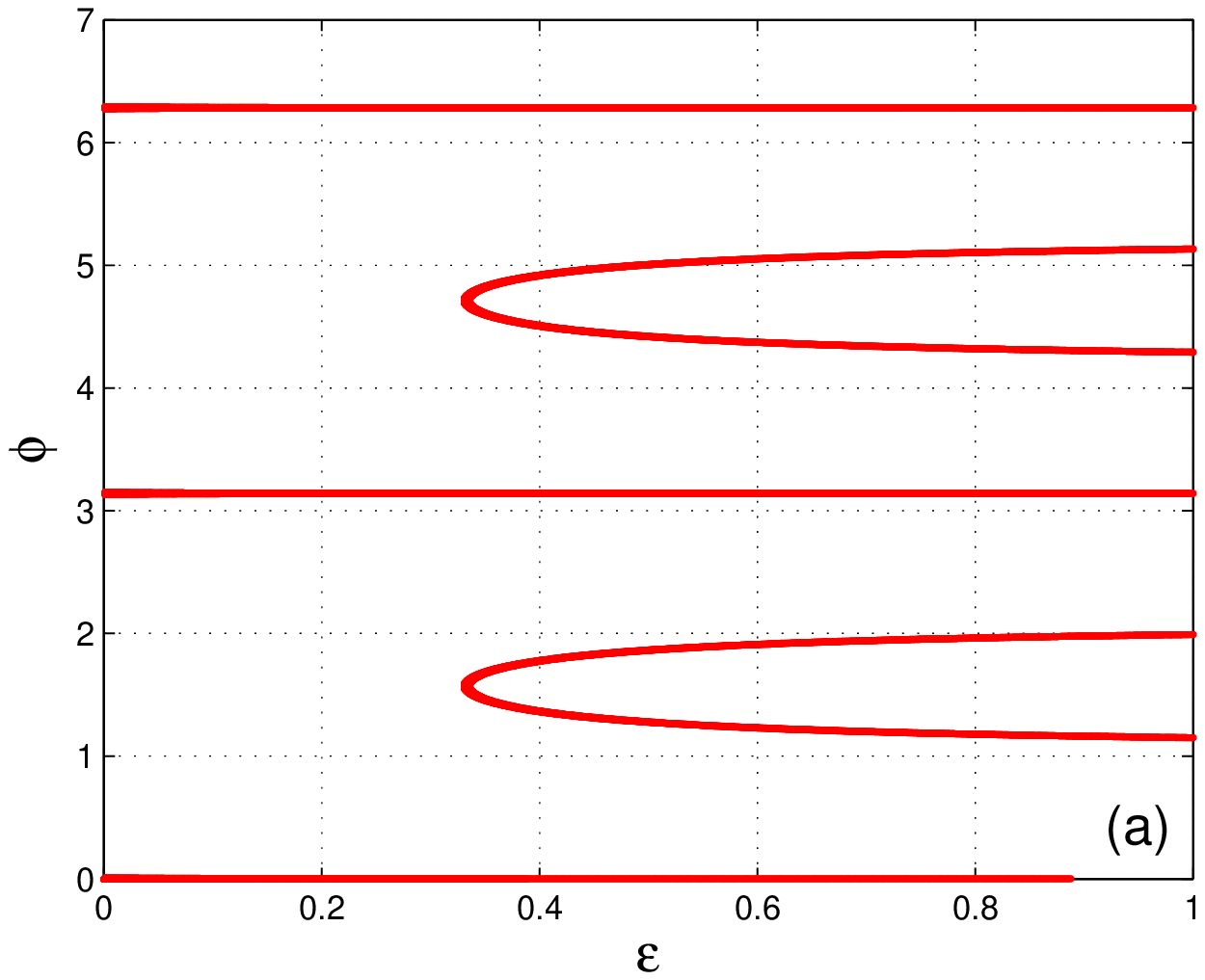}\epsfxsize=10cm\centerline{\hspace{-7cm}\epsfbox{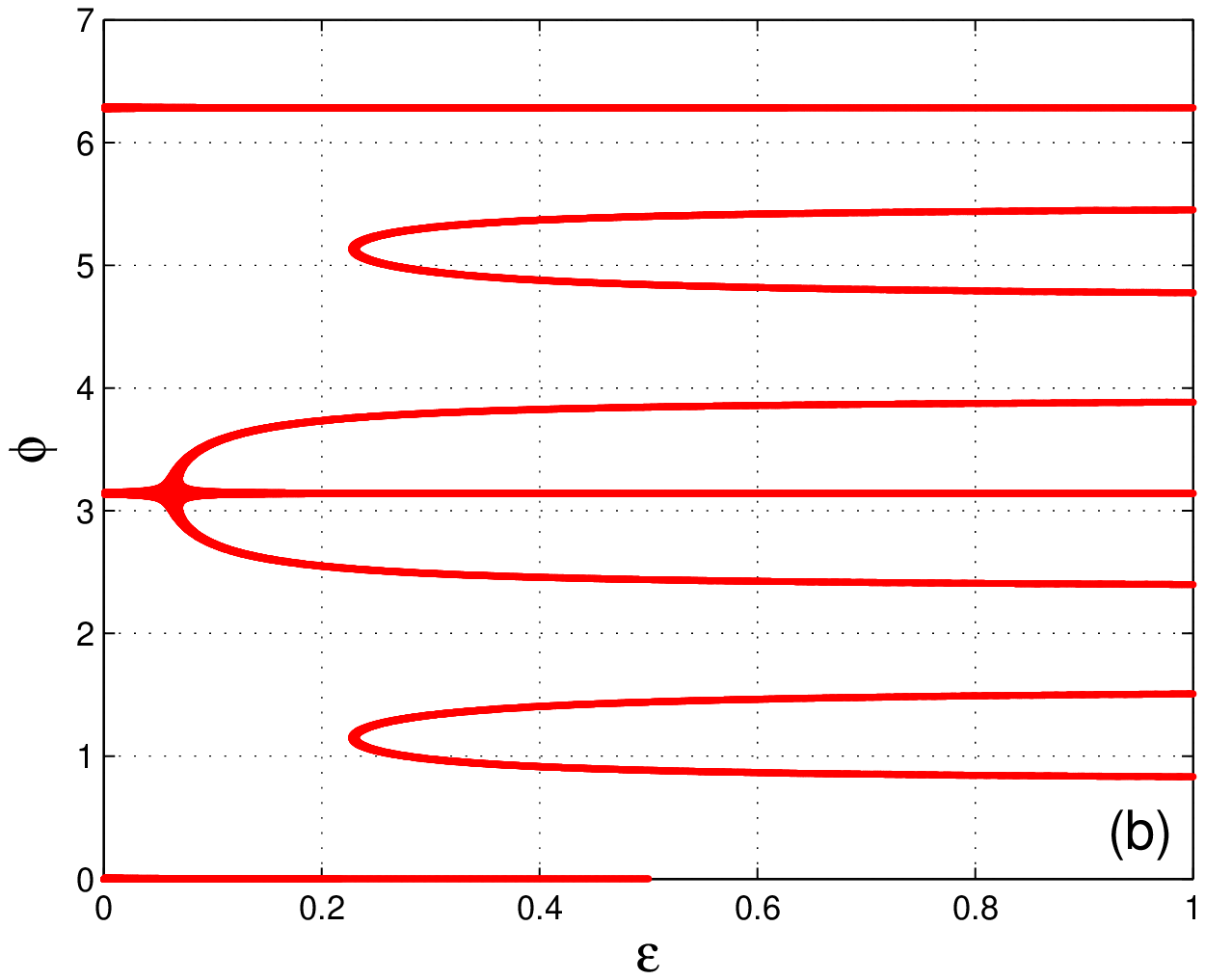}}}
\caption{Fixed points of the MSG equation for (a)$N=3$ and
(b)$N=4$.\label{12}}
\end{figure}

The fixed points of the MSG equation are obtained by equating the
RHS of the equation (\ref{shok}) to zero:
\begin{equation}
\sin \phi +N\epsilon \sin (N\phi )=0.\label{shok1}
\end{equation}
Let us call these fixed points which lie in the range $[0,2\pi]$
as $\phi_{n}(n=1,2,...)$. The number of fixed points depends on
both values of $N$ and $\epsilon$. Fig.\ref{12} shows the value
and number of fixed points as $\epsilon$ varies for two cases of
$N=3$ (Fig.\ref{12}a) and $N=4$ (Fig.\ref{12}b).

In order to investigate the stability of periodic solutions around
fixed points, we use the conventional linear perturbation
analysis, by inserting the expression
\begin{figure}[h]
\epsfxsize=8.7cm\centerline{\hspace{8cm}\epsfbox{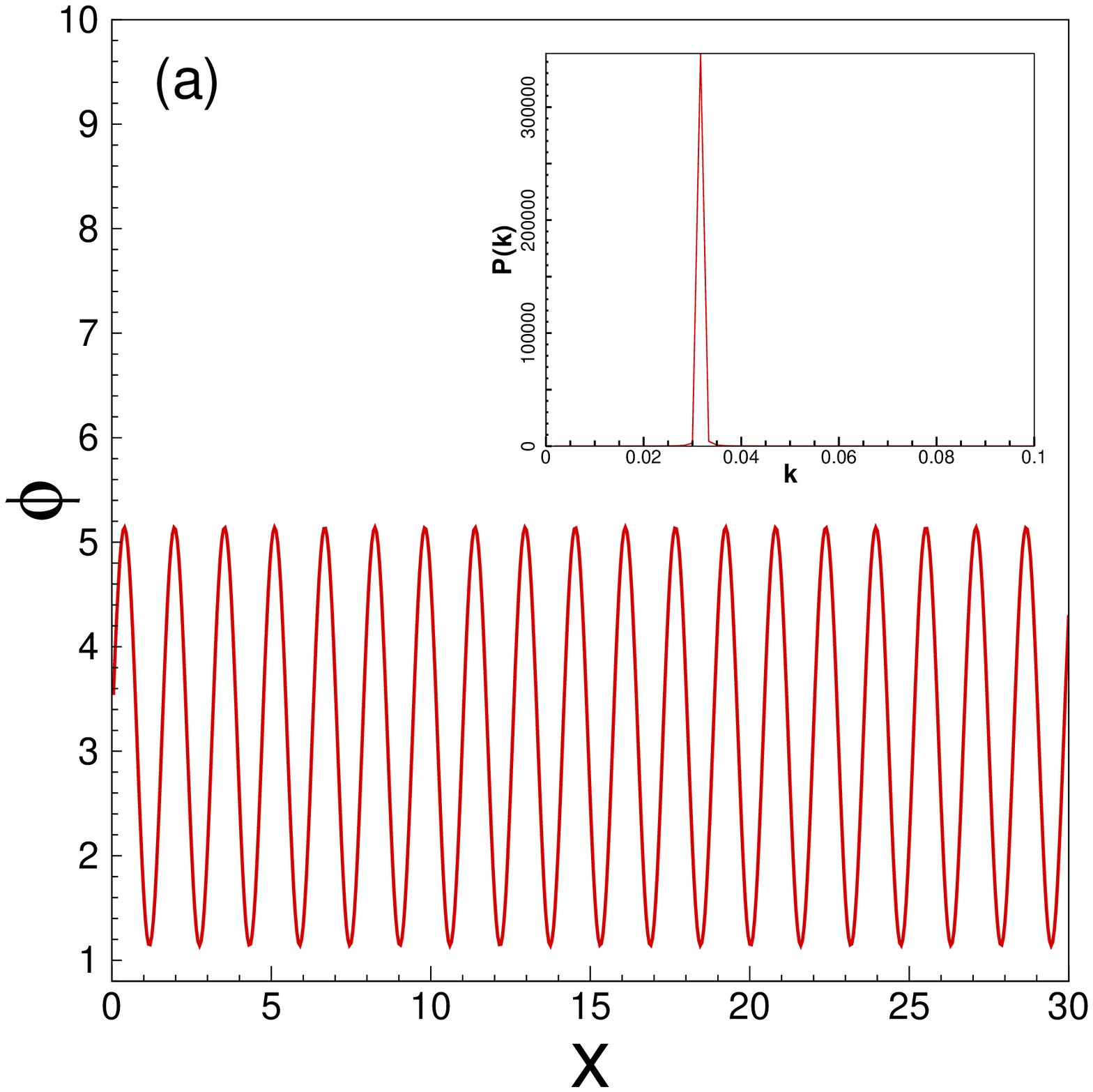}\epsfxsize=8.7cm\centerline{\hspace{-7cm}\epsfbox{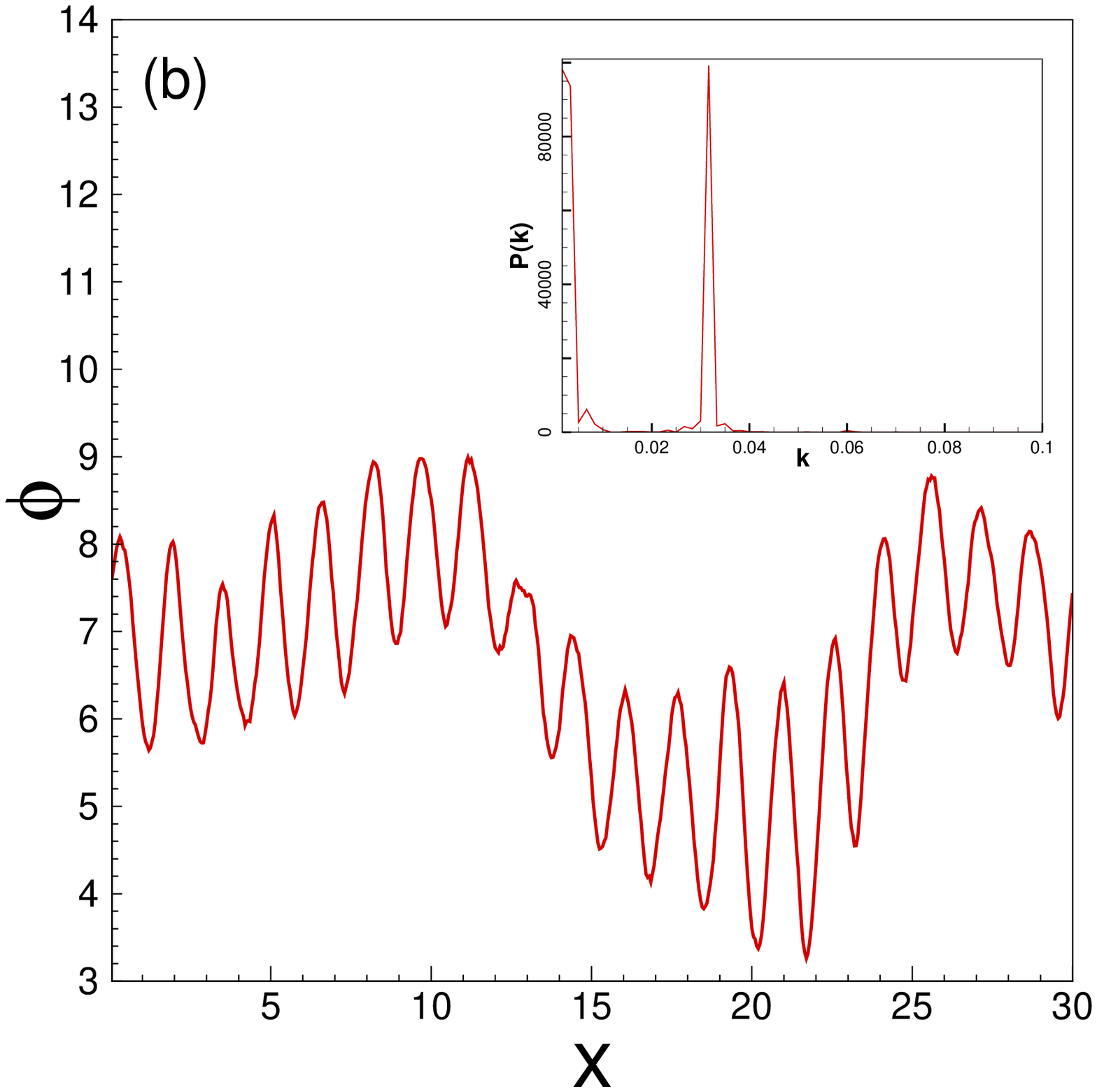}}}
\caption{(a) Initial periodic perturbation around the fixed point
of DSG ($\phi=\pi$) for $\epsilon=0.1$ and its FFT(inset), (b) The
evolved perturbation after 200 time steps and the corresponding
FFT.\label{13}}
\end{figure}
\begin{equation}
\phi(x,t)=\phi_{n} + A\cos(kx)\cos(wt),
\end{equation}
in which $A(\ll\phi_{n})$ is a constant and $k$ and $\omega$ are
wave number and angular frequency, respectively, in to the field
equation (\ref{shok}). Expanding the nonlinear sine terms and
keeping only terms linear in $A$, we obtain
\begin{equation}
\omega^{2}=k^{2}+\cos\phi_{n}+N^{2}\epsilon\cos(N\phi_{n});\label{shok2}
\end{equation}
In this dispersion relation, the stability is judged by the sign
of $\omega^{2}$ for any chosen value of $k$. Let us write
Eq.(\ref{shok2}) in the form
\begin{equation}
\omega^{2}=k^{2}+f_{n}(N,\epsilon).
\end{equation}
It is obvious that if $f_{n}>0$, $\omega^{2}$ is always greater
than zero and the perturbation is stable. If, on the other hand,
$f_{n}<0$, $\omega^{2}$ is positive and we have stability, only
for $k^{2}>-f_{n}$ or $\lambda<\frac{2\pi}{\sqrt{-f_{n}}}$. In
such a case, we expect small scale perturbations to be stable and
long ones to be unstable. The fore-mentioned linear analysis was
checked numerically, by using a finite difference
algorithm\cite{15}. We have performed numerical experiments with
periodic perturbations around a fixed point as initial condition.
In this way, we could confirm stability about the minima-energy
fixed points like $\phi_{n}=0$. Perturbations around local maxima
show a more interesting behavior. It was seen that short
wavelength modes remain stable, while long wavelength ones start
to be excited and increase in amplitude. Fig.\ref{13} shows the
fate of a periodic perturbation around $\phi_{n}=\pi$ in the DSG
system ($N=2$). The initial perturbation is shown in Fig.\ref{13}a
together with its FFT. Fig.\ref{13}b shows the perturbation after
integrating the dynamical equation for 200 time steps. As the FFT
shows, the initial, small wavelength perturbation persists, while
the unstable, long wavelength modes grow, confirming our
analytical result.

\section{concluding remarks}\label{sec6}

We have examined the periodic and step-like solutions of the MSG
system. In particular, we interpreted the step-like and periodic
solutions as a many body system of kinks and anti-kinks with a
nonlinear interaction between the solitons. This interpretation is
established by comparing the asymptotic $P\rightarrow 0$ case with
the single-soliton solution. In this limit, each site in the chain
conforms with the single-soliton solution, and the energy per
soliton approaches the single-soliton energy. As the solitons get
closer, the energy per soliton becomes different and the
difference is attributed to the interaction energy.

We observed that as $N$ increases, more and more subkinks (up to
$N$) can develop and this leads to a more complicated equation of
state. This diagram becomes multiple-valued, depending on the
number of subkinks, which in turn depends on the pressure $P$. We
examined the stability of periodic solutions around the fixed
point($\phi=\pi$ and $\frac{d\phi}{dx}=0$), obtaining the
interesting result that within the linear approximation, short
wavelength periodic solutions oscillate like standing waves
(signalling stability), while long wavelength periodic solutions
are unstable.


\begin{thebibliography}{99}
\bibitem{0} M. Peyravi,  A. Montakhab, N. Riazi and A.
Gharaati, Eur. Phys. J. {\bf B 72}, 269-277 (2009).
\bibitem{1}S. Burdick, M.El-Batanouny and C. R. Willis, Phys.
Rev. {\bf B 34}, 6575 (1986).
\bibitem{2}K. Maki and P. Kumer, Phys. Rev. {\bf B 14}, 118 (1976);
14. 3290 (1976).
\bibitem{3}Y. Shiefman and P. Kumer, Phys. Scr. 20, 435 (1979).
\bibitem{4}K. M. Leung, Phys. Rev. {\bf B 27}, 2877 (1983).
\bibitem{5}O. Hudak, J. Phys. Chem. {\bf 16}, 2641 (1983); {\bf 16}, 2659 (1983).
\bibitem{6}M. El-Batanouny, S. Burdick, K. M. Martini and P.
Stancioff, Phys. Rev. Lett. {\bf 58}, 2762 (1987).
\bibitem{7}E. Magyari, Phys. Rev. {\bf B 29}, 7082 (1984).
\bibitem{8}J. Pouget and G. A. Maugin, Phys. Rev. {\bf B 30}, 5306
(1984); {\bf 31}, 4633(1984).
\bibitem{9}N. Hatakenaka, H. Takayanagi, Y.Kasai and S. Tanda, Physica {\bf B} 284-288 (2000) 563-564.
\bibitem{10}T. Uchiyama, Phys. Rev. {\bf D 14}, 3520 (1976).
\bibitem{11}S. Duckworth, R. K. Bullough, P. J. Caudrey and J. D.
Gibbon, Phys. Lett. {\bf57 A}, 19 (1976).
\bibitem{12}V. A. Gani and A. E. Kudryavtsev, Phys. Rev. {\bf E 60}, 3305 - 3309
(1999).
\bibitem{13}M. Croitoru, J. Phys. A: Math. Gen. {\bf22}, 845-863 (1989).
\bibitem{14} C. A. Popov, Wave Motion. {\bf42(1)}, 309-350 (2006).
\bibitem{ri1} N. Riazi and A. R. Gharaati, Int. J. Theor. Phys. {\bf37},
1081 (1998).
\bibitem{15}E. Kreyszig, \textit{Advanced Engineering Mathematics}, John
Wiley and Sons, NewYork(1983).
\bibitem{16}This problem has a simple mechanical analog if the following change of variables are considred $x\longrightarrow t$, $\phi\longrightarrow x$, for a particle of unit mass.
\bibitem{17}Here, we use the word soliton interchangably with kink. The distance between successive kinks ($L$) is well-defined, and therefore energy
 per soliton ($E$) is well-defined as well.


\end{thebibliography}
\end{document}